\documentclass[letterpaper, 10 pt, conference]{ieeeconf}  % Comment this line out if you need a4paper

\IEEEoverridecommandlockouts                              % This command is only needed if 
\usepackage{circuitikz}
\usepackage{booktabs}
\usepackage{makecell}
\usepackage[nolist]{acronym}
\usepackage[colorinlistoftodos, textsize=footnotesize, textwidth=20mm]{todonotes}
\usepackage{svg} %to include svg
\usepackage{siunitx} %to get units

\usepackage{tabularx}

\usepackage{cite}
\usepackage{graphicx}
\usepackage{rotating}
\usepackage{pdflscape}

\usepackage{tikz}

\newcommand\circlearound[1]{%
  \tikz[baseline]\node[draw,shape=circle,anchor=base,scale=0.60] {#1} ;}
\usepackage{diagbox}
\usepackage{multirow}
\usepackage{array}
\newcolumntype{P}[1]{>{\centering\arraybackslash}p{#1}}
\usepackage{amssymb}
\usepackage{colortbl} % Allows table coloring

\begin{acronym}

    \acro{ro}[RO]{Remote Operator}
    \acroplural{ro}[ROs]{Remote Operators}

    \acro{av}[AV]{Automated Vehicle}
    \acroplural{av}[AVs]{Automated Vehicles}

    \acro{ads}[ADS]{Automated Driving System}

    \acro{AFGBV}[AFGBV]{Autonomous Vehicles Authorization and Operation Regulation}

    \acro{fm}[FM]{Fleet Manager}

    \acro{fo}[FO]{Field Operator}

    \acro{mrc}[MRC]{Minimal Risk Condition}

    \acro{mrm}[MRM]{Minimal Risk Maneuver}

    \acro{odd}[ODD]{Operational Design Domain}

    \acro{SAE}[SAE]{Society of Automotive Engineers}

    \acro{StVG}[StVG]{Road Traffic Act}

\end{acronym}

\title{\LARGE \bf
Control Center Framework for Teleoperation Support

of Automated Vehicles on Public Roads*
}

\author{Maria-Magdalena Wolf$^{1}$, Niklas Krauss$^{2}$, Arwed Schmidt$^{3}$ and Frank Diermeyer$^{4}$% 
\thanks{*The research was supported and funded by the Federal Ministry of Economic Affairs and Climate Action of Germany (BMWK) within the project SAFESTREAM (FKZ 01ME21007B).}%
\thanks{$^{1}$Maria-Magdalena Wolf is with Institute of Automotive Technology, Technical University of Munich, 80333            Munich, Germany
        {\tt\small maria.wolf@tum.de}}%
\thanks{$^{2}$Niklas Krauss is with Institute of Automotive Technology, Technical University of Munich, 80333                    Munich, Germany
        {\tt\small niklas.krauss@tum.de}}%
\thanks{$^{3}$Arwed Schmidt is with EasyMile GmbH, 10435 Berlin, Germany
        {\tt\small arwed.schmidt@easymile.com}}%
\thanks{$^{4}$Frank Diermeyer is with Institute of Automotive Technology, Technical University of Munich, 80333                    Munich, Germany
        {\tt\small diermeyer@tum.de}}%
}

\begin{document}

\maketitle
\thispagestyle{empty}
\pagestyle{empty}

%%%%%%%%%%%%%%%%%%%%%%%%%%%%%%%%%%%%%%%%%%%%%%%%%%%%%%%%%%%%%%%%%%%%%%%%%%%%%%%%
\begin{abstract}

Implementing a teleoperation system with its various actors and interactions is challenging and requires an overview of the necessary functions. 
This work collects all tasks that arise in a control center for an automated vehicle fleet from literature and assigns them to the two roles Remote Operator and Fleet Manager. 
Focusing on the driving-related tasks of the remote operator, a %\todo{Niklas: ist es nur ein teleoperation framework} teleoperation framework 
process is derived that contains the sequence of tasks, associated vehicle states, and transitions between the states. 
The resulting state diagram shows all remote operator actions available to effectively resolve automated vehicle disengagements. 
Thus, the state diagram can be applied to existing legislation or modified based on prohibitions of specific interactions. 
The developed control center framework and included state diagram should serve as a basis for implementing and testing remote support for automated vehicles to be validated on public roads.

\end{abstract}

%%%%%%%%%%%%%%%%%%%%%%%%%%%%%%%%%%%%%%%%%%%%%%%%%%%%%%%%%%%%%%%%%%%%%%%%%%%%%%%%
\section{INTRODUCTION}

Although remote support for \acp{av} is allowed in countries like Germany~\cite{afgbv_general}, the U.K.~\cite{uk_automated_vehicles_act} and the U.S.~\cite{DMV_California}, teleoperated vehicles remain uncommon on public roads, particularly in Europe. 
This is due to the limited maturity of \acp{av}, which struggle with certain traffic situations. Teleoperation can address this by allowing remote human assistance or control, but these systems also require further development and testing for public use.

To set up a teleoperation system, preliminary work, such as the concepts %of a control center for an \ac{av} fleet 
by Feiler et al.~\cite{feiler2020} or Gontscharow et al.~\cite{gontscharow2023}, provides an overall overview of actors and tasks in a control center for an \ac{av} fleet. 
Building on these two concepts, this work specifies the interaction process between the remote operator and \ac{av}. 
Additional role and task definitions compiled by Schwindt et al.~\cite{schwindt2023will} serve as a further basis. 
% Further research also includes role and task descriptions compiled and categorized by Schwindt et al.~\cite{schwindt2023will}. 
Expanding upon Schwindt et al.'s framework requirements~\cite{schwindt2023will}, this work integrates insights from additional literature to enhance the task categorization. 
% Building on the framework requirements by Schwindt et al.~\cite{schwindt2023will}, this work expands their task categories with insights from additional literature. 
The remote operator tasks, previously considered individually or as sequence diagrams to solve specific incidents~\cite{parr2023investigating}, are then brought into a %generic 
comprehensive state diagram that represents a generally applicable sequence and connection of the tasks in a logical process flow. %based on the vehicle's state. %taking the vehicle's state into account. 

This work provides a concrete foundation for technical implementation by defining the interactions between the remote operator and the vehicle. 
Existing systems should be evaluated against the framework to identify coverage of relevant interactions. 
Moreover, the framework offers a common base %shared communication basis 
for understanding, discussing, and further developing teleoperation. % as well as a methodological baseline for further development. 
Thereby, the focus lies on supporting Level-4 or Level-5 vehicles, defined by the \ac{SAE} J3016~\cite{SAEJ3016}, which operate autonomously unless operator intervention is triggered by a failure. 
% This assumes that the \ac{av} alerts the remote operator in case of a failure. Otherwise, it drives independently. 
Though designed for passenger transport, the framework and state diagram can be adapted to domains like logistics, where tasks such as passenger communication are unnecessary. 
% Therefore, this work provides a concrete starting point for the technical implementation by defining the interaction between the remote operator and the vehicle. 
% Existing implementations should be evaluated against the framework to identify whether all interactions are considered, or why not. The framework thus serves as a common basis for understanding and discussing teleoperation technology. 
% The work focuses on supporting Level-4 or Level-5 vehicles, as defined by the \ac{SAE} J3016~\cite{SAEJ3016}. 
% This assumes that the \ac{av} alerts the remote operator in case of a failure. Otherwise, it drives independently. 
% Though designed for passenger transport, the control center framework and state diagram can be adapted to domains like logistics, where tasks such as passenger communication are unnecessary.
% While focused on passenger transportation, the control center framework and state diagram for remote operator tasks can be adapted to other domains, such as the logistics sector, whereby tasks like passenger communication are no longer required.

\subsection{Taxonomy}

The text differentiates between the \ac{av} and \ac{ads}. 
The term \ac{av} includes the whole vehicle's hardware and software. 
In contrast, the \ac{ads} represents a subset of the \ac{av}, focusing on the vehicle's sensors, actuators, and software responsible for the automated driving task~\cite{eu-1426-2022}.
In addition, the terms \ac{mrm} and \ac{mrc} are used \cite{eu-1426-2022}: %from the Regulation (EU) 2022/1426~\cite{eu-1426-2022} are used:

\begingroup
\leftskip1em
\rightskip\leftskip
An \ac{mrm} is a maneuver intended to minimize traffic risks by safely stopping the vehicle in an \ac{mrc}.

An \ac{mrc} refers to a stable, stationary vehicle state that minimizes the crash risk.
\par
\endgroup
Thereby, the vehicle is assumed to always enter a suitable \ac{mrc} safely, eliminating the need for an emergency stop during the \ac{mrm}.

 \subsection{Contributions}

While companies like Waymo~\cite{waymo}, Zoox~\cite{zoox}, or Cruise~\cite{cruise} use teleoperation on public roads, sources on implementing such systems are scarce. 
This work defines and distinguishes the roles of the Remote Operator and Fleet Manager in an \ac{av} fleet control center, compiling their tasks in Section \ref{related_work}. 
Focusing on the driving-related tasks, a group of experts developed a generic state diagram for all remote operator interactions with \acp{av}, shown in Section \ref{framework} %, detailing interactions, vehicle states, and transitions 
to guide teleoperation system implementation.

However, with countries like Germany~\cite{afgbv_general}, the U.K.~\cite{uk_automated_vehicles_act}, and the U.S.~\cite{DMV_California} introducing legislation for \ac{av} type approval and teleoperation, specific interactions, such as remote driving in Germany, are still prohibited.
Section \ref{application} adapts the state diagram to German legislation, demonstrating its applicability and functionality as a foundational structure for remote operator support of \acp{av} on public roads.
Ultimately, the generic diagram aims to outline the remote operator's actions to resolve a high variety of \ac{av} disengagements.

%We aim to close this gap and provide a general conceptual framework that shows the different interactions, states, and transitions between states that can serve as a basis for implementing such a system. 

%%%%%%%%%%%%%%%%%%%%%%%%%%%%%%%%%%%%%%%%%%%%%%%%%%%%%%%%%%%%%%%%%%%%%%%%%%%%%%%%
\section{RELATED WORK}
\label{related_work}

\subsection{Roles in a Control Center for an AV Fleet}
%\subsection{Task groups in a Teleoperation Control Center}
\label{roles}

Operating a control center typically requires multiple individual roles. 
This section defines control center roles for an \ac{av} fleet, derived from scientific literature due to limited public information on large-scale implementations.
% This section  provides an overview of the existing roles within the control center for an \ac{av} fleet to subsequently assign tasks to these roles. Afterward, the remote operator tasks are situated within the developed state diagram. 
% Since there is no such information about a control centers for an \ac{av} fleet at scale publicly available, the roles are derived from scientific literature. 
%A review of existing publications reveals a multitude of role designations, the majority of which differ only in wording, not in definition. 
In the architecture proposed by Gontscharow et al.~\cite{gontscharow2023}, our attention is directed toward the roles of the control center for interaction with the \ac{av} fleet and primary tasks associated with transporting passengers on public roads. 
%Roles like passengers or the fire department are not described in isolation. Instead, their interfaces with the control center are defined. 
Roles like passengers or the fire department are defined through their interfaces with the control center rather than in isolation.

With regard to the various tasks in the control center, Dix et al.~\cite{dix2021autonom} argue that one person cannot simultaneously support a vehicle and oversee an entire vehicle fleet due to the cognitive demands of building situational awareness. 
Therefore, the \ac{SAE}~\cite{SAEJ3016} divides the (human) actors into passenger, dynamic driving task fallback-ready user, driverless operation dispatcher, remote assistant and (remote) driver, whereby we only consider the last three roles mentioned %in the Teleoperation Control Center 
for Level 4 \acp{av}. 
In addition, Bogdoll et al.~\cite{bogdoll2022taxonomy} summarize the terms remote driver and remote assistant in the general term remote operator. 
This leads to the differentiation between the roles of the remote operator and the dispatcher~\cite{schwindt2023will}.

This publication refers to the dispatcher as the \textit{Fleet Manager}, as the role extends beyond dispatching to include other fleet operations~\cite{SAEJ3016}, misleading the term dispatcher.
%In this publication, the dispatcher is referred as \textit{Fleet Manager} as this role encompasses more than the pure dispatching like other fleet operations functions~\cite{SAEJ3016}. Therefore, the designation dispatcher would be misleading. 
Furthermore, the terms \textit{Remote Operator} and \textit{Teleoperator} are used analogously.%, following the prevailing conventions of literature and usage.

\textbf{Fleet Manager:}
The Fleet Manager is surveilling multiple vehicles~\cite{Kalaiyarasan2021} or the whole vehicle fleet~\cite{schwindt2023will, gontscharow2023} to detect and react to abnormalities. To maintain an overview of the vehicle fleet, the Fleet Manager is responsible for coordinative and higher-level tasks such as the strategic driving task, i.e., global path planning in terms of navigation~\cite{schwindt2023will, 5GAA}. The Fleet Manager does not carry out more detailed dynamic driving tasks and does not make any safety-critical decisions~\cite{Kalaiyarasan2021}. These tasks are passed on to the Remote Operator. 

\textbf{Remote Operator:}
In contrast to the Fleet Manager, the Remote Operator interacts with only a single vehicle at a time. The Remote Operator must be able to quickly understand situations based on videos, vehicle data and maps~\cite{schwindt2023will}. Furthermore, the Remote Operator must intervene and solve problem situations by operating the vehicle from outside, beyond line-of sight while reacting to the vehicle's dynamic environment~\cite{schwindt2023will, gontscharow2023, SAEJ3016}. 
The dynamic driving task can be performed either directly by driving or indirectly by assisting the vehicle~\cite{schwindt2023will, gontscharow2023, majstorovic2022survey}.
In direct control, the Remote Operator performs both tactical and real-time operational driving tasks, i.e., behavior and local path planning in terms of path guidance and stabilization~\cite{5GAA}. In indirect control, the remote operator only performs tactical operations~\cite{5GAA}. 

Thereby, the Remote Assistant can also perform other fleet operations functions~\cite{SAEJ3016}. 
This shows that the roles overlap in their areas of responsibility and can also be assumed by one person theoretically. 
As soon as the Fleet Manager takes over the operation of a vehicle, they become a Remote Operator~\cite{SAEJ3016}. 
This is particularly conceivable in small vehicle fleets with only a few vehicles to be managed.

The described control center roles aim to replace the \textit{Safety Driver}, who is always inside the vehicle to supervise and operate as soon as the \ac{av} needs support. 
If remote support fails and on-site intervention is needed, the \textit{Field Operator}, based on Schrank and Kettwich~\cite{schrank2021}, manually operates the vehicle until the automation can take over again or removes defective vehicles. 
The following goes into more detail about the tasks in a control center to support \acp{av} and assigns these tasks to the roles of the Fleet Manager and Remote Operator.

\subsection{Tasks in a Control Center for an AV Fleet}
\label{tasks}

The tasks occurring in an \ac{av} fleet control center are analyzed based on previous literature, extending the approach of Schwindt et al.~\cite{schwindt2023will}. 
Thereby, the common terms \textit{Remote Assistance}, \textit{Remote Driving}, \textit{Remote Intervention} and \textit{Remote Monitoring} are used for driving-relevant tasks instead of Indirect and Direct Control, Release and Deactivation or Monitoring. However, we retain the task definitions proposed by Schwindt et al.~\cite{schwindt2023will} and summarize them below. 
Table \ref{tab:cc-lit} illustrates which tasks are mentioned in each source according to the following definitions. 
While the few gaps in Table~\ref{tab:cc-lit} indicate general agreement on the tasks in a control center, there is still disagreement on terminology. For example, Parr et al.~\cite{parr2023investigating} and Mutzenich et al.~\cite{Mutzenich2021} refer to Remote Assistance and Fleet Management as Remote Management. 
The gaps in the Fleet Manager’s tasks can be attributed to publications focusing on remote operator tasks. %without fully addressing or considering all tasks of the Teleoperation Control Center. 
Furthermore, the Remote Intervention is often not explicitly mentioned or implicitly included in Remote Driving or Remote Assistance.

\begin{table*}[htbp]
\renewcommand{\arraystretch}{1.2} % Hier Zeilenabstand einstellen

\caption{Overview of control-center-related literature and derived task groups assigned to the roles of the Remote Operator and Fleet Manager: White cells indicating gaps in corresponding literature}
\begin{center}
\begin{tabular}{|r|c|P{1.5cm}|P{1.5cm}|c|c|c|c|}
\cline{2-8}
\multicolumn{1}{c|}{}& \multicolumn{2}{c}{Remote Operator} & \multicolumn{4}{c}{\diagbox[rightsep=53pt, leftsep=53pt]{\quad}{\quad}} & Fleet Manager \\
\cline{2-8} 

\multicolumn{1}{c|}{} & \multicolumn{4}{c|}{\rule{0pt}{12pt}Driving-related Tasks} & \multicolumn{3}{c|}{\rule{0pt}{12pt}Mission-related Tasks (non-driving-related)} \\[5pt]
\cline{2-8}

\multicolumn{1}{c|}{} & \multirow{2}{1.5cm}{\centering \rule{0pt}{16pt}Remote Assistance} & \multicolumn{2}{c|}{\centering \rule{0pt}{16pt}Remote Control} & \multirow{2}{1.5cm}{\centering \rule{0pt}{16pt}Remote Monitoring} & \multirow{2}{*}{\centering \rule{0pt}{22pt}Communication} & 
\multirow{2}{1.5cm}{\centering \rule{0pt}{16pt}Emergency Handling} & \multirow{2}{1.5cm}{\centering \rule{0pt}{16pt}Fleet Management} \\[5pt]
\cline{3-4}

\multicolumn{1}{c|}{} & & \centering \rule{0pt}{12pt}Remote Driving & \centering \rule{0pt}{12pt}Remote Intervention & & & & \\[12pt]
\hline

Feiler et al.~\cite{feiler2020} & \cellcolor[gray]{0.7} & \cellcolor[gray]{0.7} & & \cellcolor[gray]{0.7} & \cellcolor[gray]{0.7} & \cellcolor[gray]{0.7} & \cellcolor[gray]{0.7} \\ 
\hline
Gontscharow et al.~\cite{gontscharow2023} & \cellcolor[gray]{0.7} & \cellcolor[gray]{0.7} & \cellcolor[gray]{0.7} & \cellcolor[gray]{0.7} & \cellcolor[gray]{0.7} & \cellcolor[gray]{0.7} & \cellcolor[gray]{0.7}\\ 
\hline
Schwindt et al.~\cite{schwindt2023will} & \cellcolor[gray]{0.7} & \cellcolor[gray]{0.7} & \cellcolor[gray]{0.7} & \cellcolor[gray]{0.7} & \cellcolor[gray]{0.7} & \cellcolor[gray]{0.7} & \cellcolor[gray]{0.7} \\ 
\hline
Parr et al.~\cite{parr2023investigating} & \cellcolor[gray]{0.7} & \cellcolor[gray]{0.7} & \cellcolor[gray]{0.7} & \cellcolor[gray]{0.7} & \cellcolor[gray]{0.7} & \cellcolor[gray]{0.7} & \cellcolor[gray]{0.7} \\ 
\hline
\ac{SAE} J3016~\cite{SAEJ3016} & \cellcolor[gray]{0.7} & \cellcolor[gray]{0.7} & & \cellcolor[gray]{0.7} & & & \cellcolor[gray]{0.7} \\ 
\hline
Dix et al.~\cite{dix2021autonom} & \cellcolor[gray]{0.7} & \cellcolor[gray]{0.7} & \cellcolor[gray]{0.7} & \cellcolor[gray]{0.7} & \cellcolor[gray]{0.7} & \cellcolor[gray]{0.7} & \\ 
\hline
Bogdoll et al.~\cite{bogdoll2022taxonomy} & \cellcolor[gray]{0.7} & \cellcolor[gray]{0.7} & & \cellcolor[gray]{0.7} & \cellcolor[gray]{0.7} & & \cellcolor[gray]{0.7} \\ 
\hline
Kalaiyarasan et al.~\cite{Kalaiyarasan2021} & \cellcolor[gray]{0.7} & \cellcolor[gray]{0.7} & \cellcolor[gray]{0.7} & \cellcolor[gray]{0.7} & \cellcolor[gray]{0.7} & \cellcolor[gray]{0.7} & \cellcolor[gray]{0.7} \\ 
\hline
Majstorovi{\'c} et al.~\cite{majstorovic2022survey} & \cellcolor[gray]{0.7} & \cellcolor[gray]{0.7} & \cellcolor[gray]{0.7} & \cellcolor[gray]{0.7} & & & \\ 
\hline
Schrank et Kettwich~\cite{schrank2021} & \cellcolor[gray]{0.7} & \cellcolor[gray]{0.7} & & \cellcolor[gray]{0.7} & & \cellcolor[gray]{0.7} & \cellcolor[gray]{0.7} \\ 
\hline
Mutzenich et al.~\cite{Mutzenich2021} & \cellcolor[gray]{0.7} & \cellcolor[gray]{0.7} & \cellcolor[gray]{0.7} & \cellcolor[gray]{0.7} & \cellcolor[gray]{0.7} & \cellcolor[gray]{0.7} & \cellcolor[gray]{0.7} \\ 
\hline
Pitzen et al.~\cite{pitzen2023handlungsleitfaden} & \cellcolor[gray]{0.7} & \cellcolor[gray]{0.7} & \cellcolor[gray]{0.7} & \cellcolor[gray]{0.7} & \cellcolor[gray]{0.7} & \cellcolor[gray]{0.7} & \\ 
\hline
Kettwich et al.~\cite{kettwich2021} & \cellcolor[gray]{0.7} & \cellcolor[gray]{0.7} & & \cellcolor[gray]{0.7} & \cellcolor[gray]{0.7} & & \\ 
\hline
Amador et al.~\cite{Amador2022} & \cellcolor[gray]{0.7} & \cellcolor[gray]{0.7} & & \cellcolor[gray]{0.7} & & & \\ 
\hline
Kettwich et al.~\cite{kettwich2022helping} & \cellcolor[gray]{0.7} & \cellcolor[gray]{0.7} & \cellcolor[gray]{0.7} & \cellcolor[gray]{0.7} & \cellcolor[gray]{0.7} & \cellcolor[gray]{0.7} & \cellcolor[gray]{0.7} \\ 
\hline
Tener et Lanir~\cite{Tener2023} & \cellcolor[gray]{0.7} & \cellcolor[gray]{0.7} & \cellcolor[gray]{0.7} & \cellcolor[gray]{0.7} & \cellcolor[gray]{0.7} & \cellcolor[gray]{0.7} & \cellcolor[gray]{0.7} \\ 
\hline
Biletska et al.~\cite{biletska2021} & \cellcolor[gray]{0.7} & \cellcolor[gray]{0.7} & \cellcolor[gray]{0.7} & \cellcolor[gray]{0.7} & \cellcolor[gray]{0.7} & \cellcolor[gray]{0.7} & \cellcolor[gray]{0.7}\\ 
\hline
Herzberger et al.~\cite{herzberger2022control} & \cellcolor[gray]{0.7} & \cellcolor[gray]{0.7} & \cellcolor[gray]{0.7} & \cellcolor[gray]{0.7} & \cellcolor[gray]{0.7} & \cellcolor[gray]{0.7} & \cellcolor[gray]{0.7}\\ 
\hline
Kettwich et al.~\cite{kettwich2020} & \cellcolor[gray]{0.7} & \cellcolor[gray]{0.7} & \cellcolor[gray]{0.7} & \cellcolor[gray]{0.7} & \cellcolor[gray]{0.7} & \cellcolor[gray]{0.7} & \cellcolor[gray]{0.7} \\ 
\hline
Brecht et al.~\cite{Brecht2024} & \cellcolor[gray]{0.7} & \cellcolor[gray]{0.7} & \cellcolor[gray]{0.7} & \cellcolor[gray]{0.7} & & \cellcolor[gray]{0.7} & \cellcolor[gray]{0.7} \\ 
\hline
Scoliege et al.~\cite{scoliege2022} & \cellcolor[gray]{0.7} & \cellcolor[gray]{0.7} & & \cellcolor[gray]{0.7} & \cellcolor[gray]{0.7} & \cellcolor[gray]{0.7} & \\ 
\hline

\end{tabular}
\label{tab:cc-lit}
\end{center}
\vspace{-0.1cm}
\end{table*}

%Driving-related task

% The driving-realted tasks can be split into three categories (Remote Assistance, Remote Control, Remote Monitoring) based on the level of control the remote operator has over the \ac{av}.
The driving-related tasks can be divided into \textit{Remote Assistance}, \textit{Remote Control} and \textit{Remote Monitoring} depending on the remote operator's level of control over the \ac{av}.

\textbf{Remote Assistance:}
During Remote Assistance, also known as indirect teleoperation%\cite{feiler2020, gontscharow2023, pitzen2023handlungsleitfaden, kettwich2021, schwindt2023will, Mutzenich2021, schrank2021}
, the remote operator supports the \ac{av} without directly performing the dynamic driving task~\cite{bogdoll2022taxonomy, Amador2022} by providing high-level guidance~\cite{kettwich2022helping, Tener2023}, such as offering information or advice~\cite{SAEJ3016}. % without directly performing the dynamic driving task~\cite{Amador2022}.
The remote operator can interact via abstract maneuver-based driving%\cite{biletska2021, herzberger2022control, majstorovic2022survey, kettwich2020, kettwich2022helping, schwindt2023will, dix2021autonom, parr2023investigating, Kalaiyarasan2021}
, such as a maneuver proposal or maneuver clearance. 
That means either a human suggestion is translated into driving actions~\cite{kettwich2021, pitzen2023handlungsleitfaden} or a possible maneuver proposed by the \ac{av} is selected, rejected, or approved by the remote operator and executed by the \ac{av}~\cite{herzberger2022control}. 
Moreover, the remote operator can support the \ac{av} by classifying unknown objects~\cite{herzberger2022control, majstorovic2022survey}.
In general, the system is responsible for performing the driving task and the remote operator only provides support in exceptional cases~\cite{majstorovic2022survey, Brecht2024, SAEJ3016}. 
This can occur reactively at the request of the \ac{av}~\cite{scoliege2022, Kalaiyarasan2021} or proactively based on vehicle monitoring, whereby the first case should be preferred~\cite{Kalaiyarasan2021}. 
% This requires the vehicle to automatically recognize when it needs support
%, as described in R{\"o}gnvaldsson et al.~\cite{Rognvaldsson2018}, 
% to alert the remote operator with optimal situational information, eliminating the need for time-intensive and demanding permanent vehicle monitoring. 
This allows the vehicle to receive assistance within its \ac{odd}~\cite{Amador2022} or temporarily extend it, like crossing a solid line~\cite{dix2021autonom, kettwich2022helping}, with remote operator monitoring required when exiting the \ac{odd}.

During \textbf{Remote Control} the remote operator is required to continuously supervise an \ac{av}'s operation and perform a safety-critical role who can intervene in the vehicle's functions. This may include actions such as fully performing the dynamic driving task (Remote Driving) or pressing an emergency stop button (Remote Intervention)~\cite{Kalaiyarasan2021}.

\textbf{Remote Driving:}
During Remote Driving, also known as direct teleoperation%~\cite{feiler2020, gontscharow2023, Mutzenich2021, Kalaiyarasan2021, schrank2021, schwindt2023will, pitzen2023handlungsleitfaden, kettwich2021}
, the remote operator performs in real-time some or all of the dynamic driving task %\cite{SAEJ3016, bogdoll2022taxonomy, Amador2022, Mutzenich2021} 
by providing low-level longitudinal and lateral motion control~\cite{kettwich2022helping, SAEJ3016} and directly transmitting vehicle commands while having full manual control over the vehicle. %\cite{biletska2021, herzberger2022control, kettwich2020, schwindt2023will, majstorovic2022survey, Tener2023, Amador2022, Brecht2024}. 
Consequently, the remote operator is responsible for the dynamic driving task~\cite{Mutzenich2021} and can overrule the \ac{ads}~\cite{SAEJ3016}. 
This requires the remote operator to continuously monitor the environment and respond to it in a closed loop control~\cite{Kalaiyarasan2021, kettwich2021}. 
Therefore, the remote operator can proactively~\cite{scoliege2022} expand the \ac{odd} of the \ac{av}~\cite{Mutzenich2021} or take over an \ac{av}'s request if no \ac{ads} functions are available~\cite{bogdoll2022taxonomy}. %(\ac{SAE} Level 0-2)~\cite{bogdoll2022taxonomy}.

\textbf{Remote Intervention:}
Remote Intervention allows the remote operator to directly influence the movement, status or conspicuity of the \ac{av}~\cite{Kalaiyarasan2021}. This involves 
%the lowest level of vehicle control~\cite{schwindt2023will, majstorovic2022survey},
overriding the \ac{av} without cooperation or assistance, and excludes driving.
The remote operator can either release the \ac{av} to drive automatically~\cite{biletska2021, schwindt2023will} or to continue its journey after a detected situation~\cite{kettwich2020, kettwich2022helping}. On the other hand, the remote operator can bring the vehicle to a safe halt or minimal risk state %\cite{gontscharow2023, biletska2021, herzberger2022control, pitzen2023handlungsleitfaden, schwindt2023will, parr2023investigating, Tener2023}
to deactivate~\cite{pitzen2023handlungsleitfaden, schwindt2023will, dix2021autonom} or turn the vehicle off if necessary~\cite{gontscharow2023}.
To enhance conspicuity, the remote operator can use the horn %, gesture, e.g. via an external display on the vehicle, 
or lights~\cite{parr2023investigating}. 
As these interventions do not require longer interaction times and can immediately put an \ac{av} to a safe halt, this type of task could also be performed by the fleet manager. 
However, it is uncertain whether situational awareness from monitoring the entire fleet is sufficient for this.
In contrast, the operation of functions that are necessary for the driving task, such as indicators~\cite{Brecht2024, Kalaiyarasan2021}, are assigned to Remote Driving handled by the remote operator.

% \vspace{-0.06cm}
% \vspace{-\baselineskip}

\textbf{Remote Monitoring:}
Remote Monitoring refers to the surveillance of an \ac{av} by checking its status%~\cite{feiler2020, pitzen2023handlungsleitfaden, kettwich2021, kettwich2022helping, dix2021autonom, Amador2022, Brecht2024, Kalaiyarasan2021}
, traffic surroundings%~\cite{biletska2021, herzberger2022control, kettwich2020, kettwich2021, kettwich2022helping, dix2021autonom, parr2023investigating, Brecht2024, Kalaiyarasan2021}
, and vehicle interior %\cite{herzberger2022control, kettwich2020, kettwich2022helping, parr2023investigating} 
without simultaneously performing other tasks, but make safety-critical decisions, %\cite{bogdoll2022taxonomy, Amador2022, Kalaiyarasan2021}. 
in compliance with local data protection regulations. 
This allows reviewing the reason for a minimal risk condition~\cite{Mutzenich2021} or supporting if necessary~\cite{gontscharow2023, parr2023investigating, Kalaiyarasan2021}. When support is needed, Remote Monitoring transitions to Remote Assistance, Remote Driving, or Remote Intervention~\cite{Kalaiyarasan2021}. 
As detailed in Section \ref{roles}, this work distinguishes the Remote Operator, who monitors an individual vehicle, from the Fleet Manager, responsible for fleet monitoring and operational health.
%As described in Section \ref{roles}, this work differentiates between the roles of Remote Operator and Fleet Manager. The task of the remote operator is to monitor an individual vehicle, while the fleet manager is responsible for fleet monitoring, which includes surveillance of the operational health of all fleet vehicles. %\cite{feiler2020, gontscharow2023, herzberger2022control, kettwich2020, schwindt2023will, schrank2021}. 
The fleet manager oversees fleet behavior and regular operations regarding traffic or radio information~\cite{Mutzenich2021} and schedule delays~\cite{biletska2021, schrank2021}. 
Additionally, the fleet manager monitors the infrastructure %\cite{gontscharow2023, biletska2021, kettwich2020, schrank2021} 
and manages vehicle error messages 
%\cite{feiler2020, gontscharow2023, herzberger2022control, kettwich2020, schwindt2023will, schrank2021, Kalaiyarasan2021} 
to delegate driving or assisting tasks to the remote operator~\cite{schrank2021}.

%Mission-related tasks (non-driving-related):

The mission-related tasks \textit{Communication}, \textit{Emergency Handling} and \textit{Fleet Management} are primarily handled by the fleet manager. 
% Following Schwindt et al.~\cite{schwindt2023will}, the Communication category is retained, while Emergency Handling is added to encompass Other Tasks defined by Schwindt et al.~\cite{schwindt2023will}. 
% Emergency-related communication is reassigned to Emergency Handling, and the Coordination category is merged into Fleet Management, which is addressed broadly, as the state diagram focuses on the Remote Operator. 
Following Schwindt et al.~\cite{schwindt2023will}, the \textit{Communication} category is retained, with \textit{Emergency Handling} added to cover additional tasks. Emergency-related communication is moved to \textit{Emergency Handling}, and Coordination is integrated into \textit{Fleet Management}, which is broadly addressed as the state diagram focuses on the Remote Operator.

\textbf{Communication:}
To be able to exchange with the environment, Communication with 
%\vspace{-0.1cm}
\begin{itemize}
    \item passengers, e.g., to inform them about delays%~\cite{feiler2020, gontscharow2023, biletska2021, herzberger2022control, pitzen2023handlungsleitfaden, kettwich2020, kettwich2022helping, schwindt2023will, dix2021autonom, parr2023investigating, bogdoll2022taxonomy, Tener2023, Mutzenich2021, Kalaiyarasan2021},
    \item other road users, e.g., to give right of way %to another vehicle~\cite{feiler2020, kettwich2020, kettwich2022helping, schwindt2023will, parr2023investigating, Tener2023},
    \item executive forces, e.g., police officer checking the vehicle%~\cite{feiler2020, kettwich2020, schwindt2023will}, 
    \item emergency services, e.g., fire department regulating traffic%~\cite{kettwich2020, schwindt2023will, parr2023investigating, Tener2023},
    \item other control centers, e.g., to communicate with the fleet manager, or%~\cite{gontscharow2023, kettwich2020, Tener2023}, or
    \item  field operators, e.g., to ask them for help on-site%~\cite{schwindt2023will, Tener2023, Mutzenich2021}, 
\end{itemize}
must be ensured. %\todo[inline]{Wir haben field operator voher nicht erklaert, unter emergency handling wieder relevant}
Thereby, Communication must be possible in both directions from and to the remote operator or fleet manager.
Communication could be structured with the fleet manager handling general inquiries and forwarding driving-related requests to a remote operator when needed. A dedicated customer support team could also manage general inquiries, while remote operators should maintain the ability to communicate directly when interacting with the \ac{av}.

\textbf{Emergency Handling:}
In addition to Communication, Emergency Handling is distinguished as the extra task of responding immediately in emergency situations~\cite{feiler2020}. 
Incident or accident management 
%\cite{scoliege2022} 
is not limited to events inside the vehicle, such as %disputes between passengers, 
vandalism, or medical emergencies, but also includes the vehicle's surroundings, e.g., medical services for other road users~\cite{pitzen2023handlungsleitfaden, herzberger2022control} or cyberattacks. 
Therefore, the fleet manager must accept and answer emergency calls from inside or outside the shuttles~\cite{kettwich2020, schrank2021}. % or, if applicable, surrounding road users~\cite{kettwich2020, schrank2021}. 
If necessary, the fleet manager must immediately contact the relevant emergency personnel 
%\cite{schwindt2023will, parr2023investigating, Kalaiyarasan2021, kettwich2022helping} 
and initiate measures in the shuttle interior 
%\cite{kettwich2020} 
and for road safety, e.g., activating hazard lights.
%\cite{schwindt2023will}. 
When the emergency services arrive, they are to be supported remotely, e.g., via field operations~\cite{kettwich2020, schrank2021} or by contacting a remote operator to control the vehicle. 
% Once the emergency has been resolved, the field operations may have to be changed again~\cite{kettwich2020}, and a replacement vehicle may be procured~\cite{kettwich2022helping}. 
% Finally, the incident must be analyzed and documented~\cite{kettwich2021, schrank2021}. 
Once the emergency is resolved, field operations may need adjustments~\cite{kettwich2020}, a replacement vehicle may be arranged~\cite{kettwich2022helping}, and the incident analyzed and documented~\cite{kettwich2021, schrank2021}. 
If an accident occurs while the remote operator is connected, they should report it to the control center or fleet manager~\cite{Tener2023}, make an emergency call if needed, and document the incident.
% An accident can also occur while the remote operator is connected to a vehicle. In this case, the remote operator should report the incident to the control center or fleet manager~\cite{Tener2023} and, if necessary, make the emergency call independently and document the accident. 
In the event of an evacuation, both the remote operator and the fleet manager should be able to open and close the vehicle doors.%~\cite{kettwich2022helping, Brecht2024, Mutzenich2021, Kalaiyarasan2021}.
%In addition, measures must be initiated in response to abnormal incidents, such as those resulting from cyberattacks. 

\textbf{Fleet Management:}
Besides fleet monitoring, other tasks related to fleet organization are summarized under the term Fleet Management. 
This includes fleet mission planning and rerouting, also known as dispatching, to adapt operations to traffic by providing strategic instructions, such as adjusting maps sent to vehicles~\cite{feiler2020, gontscharow2023, Tener2023}.
% On the one hand, this includes fleet mission planning and modifying%~\cite{feiler2020, Kalaiyarasan2021}
% , also known as dispatching %~\cite{biletska2021, parr2023investigating, bogdoll2022taxonomy, Mutzenich2021}
% or rerouting%~\cite{gontscharow2023, herzberger2022control, kettwich2022helping, schwindt2023will, Brecht2024, schrank2021}
% , to provide strategic instructions regarding selection of destinations
% %\cite{SAEJ3016} 
% and adapt the operation plan due to traffic situations.
% %\cite{kettwich2020}. 
% Rerouting can be done, for example, by adjusting a map that is forwarded to the vehicles~\cite{feiler2020, gontscharow2023, Tener2023}. 
Furthermore, the fleet manager oversees fleet scheduling, ensuring vehicles are at the right place and time by managing trip initiation. 
While this task is mainly automated~\cite{feiler2020}, the fleet manager primarily monitors schedule adherence~\cite{kettwich2020, schrank2021}.
% On the other hand, the fleet manager is responsible for fleet scheduling, i.e., organizing the routes and vehicles in a fleet 
% %\cite{feiler2020, gontscharow2023, biletska2021, schwindt2023will} 
% to ensure that sufficient vehicles are at the right place at the right time. %~\cite{herzberger2022control}. 
% The fleet manager thus puts the vehicles into or out of service by trip initiation timing. %\cite{SAEJ3016,  biletska2021}. 
% According to Feiler et al.~\cite{feiler2020}, this task is already mainly automated, so the fleet manager only has to monitor schedule adherence~\cite{kettwich2020, schrank2021}. 
The fleet manager should also monitor the vehicles' charge status and send them for recharging if necessary. %\cite{feiler2020, biletska2021, kettwich2020, Tener2023, schrank2021}.
In addition, the fleet manager handles operational disruptions, such as accidents, by identifying, %classifying, 
and prioritizing vehicle requests. Driving and assistance tasks are forwarded to the remote operator if needed~\cite{schrank2021}. 
Moreover, monitoring and controlling infrastructure (e.g., traffic light prioritization) are possible \cite{biletska2021}.

Table \ref{tab:cc-lit} outlines the task division between the Remote Operator and Fleet Manager. 
The remote operator handles \textit{Remote Assistance}, \textit{Remote Driving}, and most \textit{Remote Intervention}, though the fleet manager can intervene if needed. \textit{Remote Monitoring} is shared, with the remote operator focusing on individual vehicles and the fleet manager overseeing the fleet. Mission-related tasks are mainly handled by the fleet manager, but \textit{Communication} and \textit{Emergency Handling} are also accessible to the remote operator. \textit{Fleet Management} remains the fleet manager's responsibility.

While some publications mention all tasks in a control center for an \ac{av} fleet, none have structured them into a process to derive a state diagram.

%%%%%%%%%%%%%%%%%%%%%%%%%%%%%%%%%%%%%%%%%%%%%%%%%%%%%%%%%

\section{METHODOLOGY}

%To complete the tasks collected in the literature and 
To develop a generally applicable control center framework for an \ac{av} fleet as well as a state diagram 
for the remote operator interactions with the \ac{av}, 
experts from teleoperation, \ac{ads} manufacturing, \ac{av} fleet operations, mobile communications, and \ac{av} testing %collaborated regularly to share their insights. 
% a consortium of teleoperation experts, an \ac{ads} manufacturer, operators of \ac{av} fleets, a mobile communications company and an \ac{av} testing organization 
exchanged experience from different perspectives at regular meetings.

Initially, the tasks of safety drivers in \acp{av} were recorded through training documents, interviews, and route visits at two locations. 
However, since safety drivers operate Level-2 vehicles, many tasks, such as clearing junctions or crosswalks, are expected to be handled independently by Level-4 vehicles. 
Additionally, remote operators, physically separated from the vehicle, would no longer need to monitor \acp{av} continuously but only provide support in exceptional cases.
% In the first step, the tasks of the safety drivers currently operating in \acp{av} were recorded based on training documents, semi-structured interviews, and visits to the routes at two different locations. 
% However, it had to be noted that the safety drivers operate in Level-2 vehicles and take on tasks that can be assumed to be carried out independently by a Level-4 vehicle due to the increasing degree of automation, e.g., constant clearance to continue driving at junctions or crosswalks should no longer be required. 
% In addition, the remote operator, who is physically separated from the vehicle, should no longer have to continuously monitor the \ac{av} but only provide support in exceptional cases. 
Therefore, generic tasks of the remote operator, which are independent of location and situation, had to be derived in regular discussions 
%using the expertise and experience of the consortium 
using literature as a basis. 
These tasks were placed in a logical sequence as an overall context in the form of a state diagram. 
Moreover, the resulting generic control center framework for an \ac{av} fleet, including the state diagram, aims to be compliant 
%was tested for feasibility in accordance 
with the German legislation and was deemed to be feasible by the experts by deriving a specific \ac{AFGBV}~\cite{afgbv_general}, \ac{StVG}~\cite{stvg} and Regulation (EU) 2022/1426~\cite{eu-1426-2022} conformable solution. 

The following section outlines the remote operator's tasks and their connections in the associated state diagram. 
% The tasks of the remote operator and their connection in the associated state diagram are described and presented in the following section.

%%%%%%%%%%%%%%%%%%%%%%%%%%%%%%%%%%%%%%%%%%%%%%%%%%%%%%%%%%%%%%%%%%%%%%%%%%

\section{RESULTS}

\subsection{State Diagram for the Remote Operator Tasks}
\label{framework}

%The main tasks of the remote operator support an \ac{av} in carrying out the dynamic driving task. 
%These driving-related teleoperation tasks include all tasks required to maneuver a road vehicle, whereby the remote operator can only operate one vehicle at a time. 
To outline a typical daily \ac{av} service procedure, the \textit{Remote Monitoring} task from section \ref{tasks} is divided into \circlearound{1}~\textit{Start/End of Service}, which describes the initial remote operator-\ac{av} connection, and \circlearound{4}~\textit{Start/End of Monitoring} during operation. \textit{Remote Intervention} is further subdivided into \circlearound{2}~\textit{De-/Activation}, \circlearound{3}~\textit{Automation Engagement} and \circlearound{5}~\textit{Minimal Risk Maneuver}, while \textit{Remote Assistance} and \textit{Remote Driving} fall under the super-category \circlearound{6}~\textit{Alternative Maneuver}. 
%Additional functions like are part of driving-related duties but not detailed here. 
Mission-related tasks, such as \textit{Emergency Handling}, and functions like operating the horn, or wipers are not considered as they could occur continuously during the remote operator's driving-related interactions. 
\textit{Fleet Management} tasks handled by the fleet manager are also excluded.

To clarify the defined sub-tasks, we outline the following procedure, covering the start to the end of an \ac{av} service, visualized in the accompanying state diagram (Fig. \ref{fig:TeleopFramework}). %as part of a remote operator's routine. 

At the start of the service, the \ac{av} is parked at a safe location in an \textit{Initial State}. 
Before starting the service, a \ac{fo} must prepare the vehicle on-site by unplugging it from the charging station, checking its roadworthiness, switching on the \ac{av}'s sensors and actuators, and ensuring that the \ac{av} is in an \ac{mrc}. However, the \ac{ads} is still deactivated and in the \textit{Prepared State}.

\circlearound{1} \textit{Start of Service:}
As it is assumed that the \ac{ro} is intended to be part of the system and service, the connection to the vehicle must be tested before starting the automated driving. To test the prepared vehicle's connection with the \ac{ro}, the \ac{fm} informs the \ac{ro} that the vehicle has been prepared. To launch the service, the \ac{ro} connects to the \ac{av} and starts the interaction so that they can monitor the vehicle from remote. 
Alternatively, it is conceivable that the vehicle automatically registers in the teleoperation system once the sensors and actuators have been started and goes into a \textit{Deactivated \ac{mrc} State}.

\circlearound{2} \textit{Activation:}
The \ac{av} is now in interaction with the \ac{ro} (or the teleoperation system) and all requirements for automated driving are met so that the \ac{av} is prepared (e.g., vehicle is roadworthy, sensor check was successful and actuators are ready), but the \ac{ads} is still deactivated. The \ac{ro} activates the \ac{ads}. 
If the secure connection setup and video transmission can be captured by the teleoperation system itself in the future, the system could also activate the \ac{ads} after registration. If the activation is successful, the \ac{ads} starts calculating a starting trajectory in \textit{Activated \ac{mrc} State} and is ready for automated driving. However, the vehicle waits for the \ac{ro} to authorize the start of the automated drive. 

\circlearound{3} \textit{Automation Engagement:}
The \ac{av} is activated in the \ac{mrc}, and the \ac{ro} is in interaction. 
Provided an automated continuation on the route is possible and there is no remaining reason for the \ac{mrc}, the \ac{ro} gives the go-ahead for the \ac{av} to leave the \ac{mrc} and start or continue in a \textit{Monitored Automated Driving State}. 
% todo[inline]{NK: and releases the MRC?}
The maneuver continues on the path calculated by the vehicle.
%on the actual lane and corresponds to a part of the planned route.

\circlearound{4} \textit{End of Monitoring:}
The \ac{ro} monitors the vehicle movement during the automated drive and can end the interaction with the \ac{av} at any time. 

The \ac{av} then drives independently in an \textit{Unmonitored Automated Driving State}. % automated mode without being monitored by the \ac{ro}.
This state represents the desired condition in which the entire vehicle fleet is only supervised by the \ac{fm}. 
Thereby, the \ac{ads} continuously transmits data on the vehicle status to the control center (e.g., GPS position). %(e.g., current speed, loading status, GPS position).

Up to this point, regular operation has been described. 
In the following, it is assumed that there is an indication that a malfunction or failure of the \ac{ads} could occur.

%(e.g., passengers report an \ac{av} inconsistency to the \ac{fm} or the \ac{fm} recognizes a possible critical vehicle condition that has not yet been reported by the \ac{av})

\circlearound{4} \textit{Start of Monitoring:}
If the \ac{fm} determines the need for further vehicle examination or intervention (e.g., due to unexpected behavior or discrepancies in vehicle values without being reported by the \ac{av}), they request the \ac{ro}. 
The \ac{ads} may also ask the \ac{ro} to monitor the system in complex, uncertain situations.
These cases result in the \ac{ro} starting interaction and thus the \textit{Monitored Automated Driving State}. %monitoring the \ac{av}.

\circlearound{5} \textit{Minimal Risk Maneuver:}
The \ac{av} moves automatically while being monitored by the \ac{ro}. 
If the \ac{ro} or \ac{ads} deems it necessary, an \ac{mrm} can be triggered.
%Alternatively, if the \ac{ro} is not yet in interaction, the \ac{ads} can trigger an \ac{mrm} and request the \ac{ro} in parallel, who then starts interaction. 
If the \ac{ads} triggers an \ac{mrm} without being monitored by the \ac{ro}, it automatically sends an interaction request to the \ac{ro}. 
The \ac{av} subsequently performs the \ac{mrm} and goes into an \textit{Activated \ac{mrc} State}.

Afterward, the autonomy must either be engaged (\textit{Automation Engagement}), deactivated (\textit{Deactivation}), or an \textit{Alternative Maneuver} be entered. 
If it is not possible for the \ac{av} to continue the route automatically, an \textit{Alternative Maneuver} must be entered, or \textit{Deactivation} may be necessary.

\circlearound{6} \textit{Alternative Maneuver:}
The \ac{av} is in the \textit{Activated \ac{mrc} State}, monitored by the \ac{ro}. 
As a continuation on the route is not possible (e.g., an object is blocking the roadway and the \ac{av} cannot calculate a valid trajectory), the \ac{ro} must define an alternative maneuver. 
The maneuver can be entered via various interaction options, so-called teleoperation concepts~\cite{majstorovic2022survey}. 
A distinction is made between \textit{Remote Driving} concepts, in which the \ac{ro} takes full and direct control of the vehicle, and \textit{Remote Assistance} concepts, in which the \ac{ro} cooperates with the \ac{av} and performs the dynamic driving task together through indirect control like described in Section \ref{tasks}.
The teleoperation concept can either be automatically suggested by the system or selected by the \ac{ro}, where different scenarios require different teleoperation concepts~\cite{Brecht2024}. 
If no alternative maneuver can be specified using a teleoperation concept and therefore the attempt fails, e.g., due to a lack of available \ac{ads} functions or bad network quality, the vehicle remains in the \textit{Activated \ac{mrc} State}.
If the maneuver is successfully entered, the \ac{av} can continue to drive independently in the \textit{Monitored Automated Driving State} and the \ac{ro} can monitor the vehicle until they end the interaction with the \ac{av} at some point to go into the \textit{Unmonitored Automated Driving State} as described in \circlearound{4}.

\circlearound{2} \textit{Deactivation:}
%The \ac{ro} is interacting with the \ac{av} in \ac{mrc} and assesses the situation. 
Since neither the automated continuation of the journey nor an alternative maneuver is possible (e.g., due to a technical defect or an accident) or is planned (e.g., at the end of the day), the \ac{ro} initiates deactivation of the \ac{ads} to stop the \ac{ads} calculating a trajectory and switch to the \textit{Deactivated \ac{mrc} State}.

\begin{landscape}
\vspace{-0.4cm}
\begin{figure}[p]
	\centering
    \resizebox{\columnwidth}{!}{
		\includegraphics[scale=0.09]{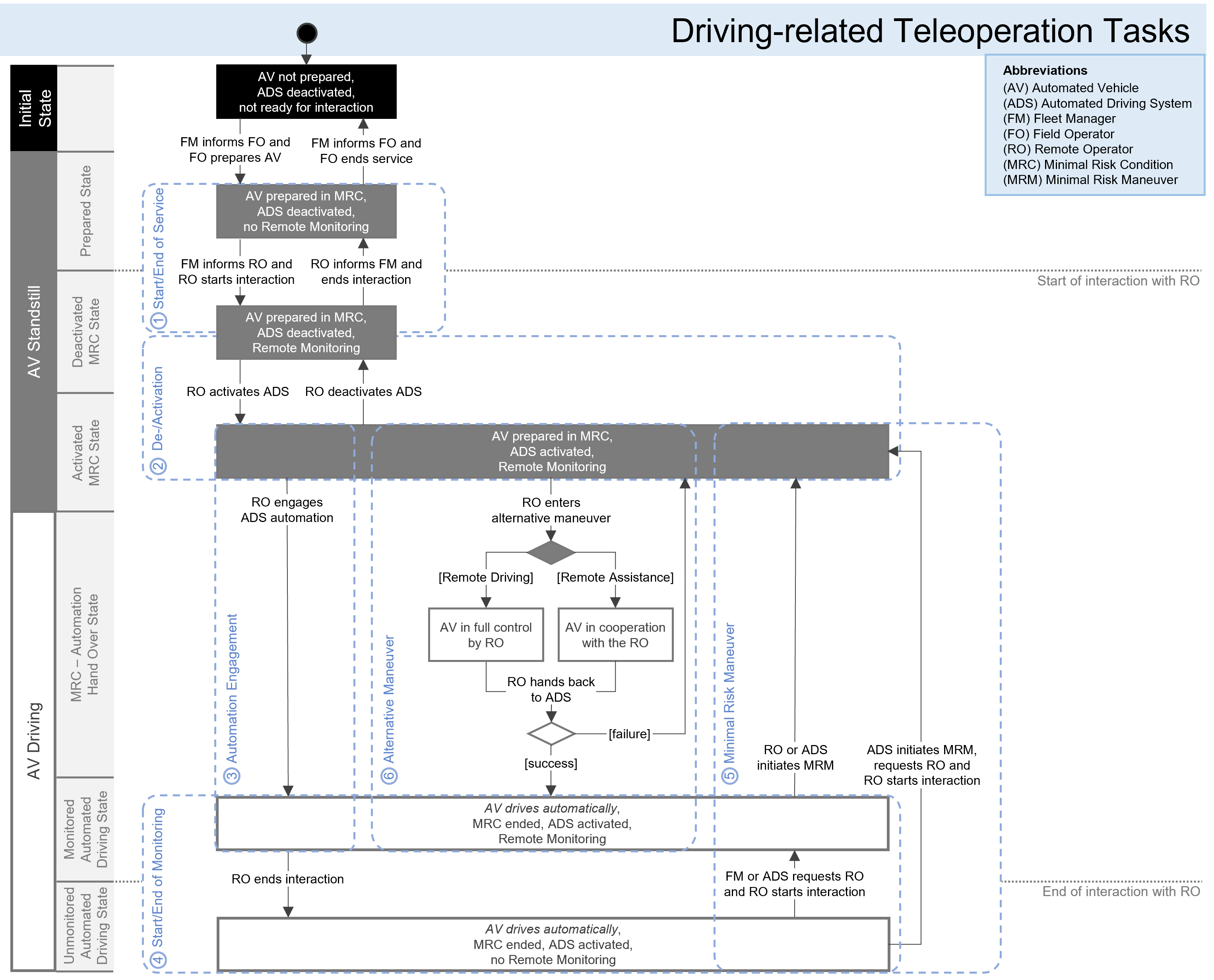}
	}
    \vspace{-0.6cm}
	\caption{State diagram organizing the remote operator's driving-related tasks \protect\circlearound{1}-\protect\circlearound{6} into a logical sequence for an assumed daily \ac{av} service procedure with starting point in the upper black circle \\ to the target state of unmonitored automated driving at the bottom} 
	\label{fig:TeleopFramework}
\end{figure}
\end{landscape}

\circlearound{1} \textit{End of Service:}
To terminate the service, the \ac{av} is in a \textit{Deactivated \ac{mrc} State}%and the \ac{ads} has been deactivated by the \ac{ro}
, e.g., because no automated further travel is to take place at the end of the day or no further travel is possible due to a technical defect or an accident. The \ac{ro} informs the \ac{fm} about the end of the \ac{av} service.
Thereupon, the \ac{fm} informs the \ac{fo}, who goes to the location of the \ac{av}, if necessary removes the \ac{av} from the road traffic and ends the driving operation (switching off sensors and actuators, charging or similar) to set the \ac{av} in \textit{Initial State}.

For future systems, it is conceivable that \circlearound{2} \textit{Deactivation} automatically accompanies the \circlearound{1} \textit{End of Service}, i.e. the information to the \ac{fm} and, if necessary, the deployment of the \ac{fo}. This minimizes and shifts the start of the interaction with the \ac{ro} in the state diagram in Fig. \ref{fig:TeleopFramework} downwards.

\subsection{Adjustment of the State Diagram to German Legislation}
\label{application}

To assess the applicability of the developed control center framework, the state diagram is applied to German law~\cite{afgbv_general, stvg, eu-1426-2022}.
%law in force in Germany by the \ac{AFGBV}~\cite{afgbv_general}, \ac{StVG}~\cite{stvg} and the Regulation (EU) 2022/1426~\cite{eu-1426-2022}. 
Thereby, remote operator tasks were derived from legislation and compared with the defined tasks in Section \ref{framework}. %, see Table \ref{TabTasksLegislation}.
As shown in Table \ref{TabTasksLegislation}, German law currently prohibits \textit{Remote Driving},
%directly driving a vehicle from remote, 
making this %the \textit{Remote Driving} 
path during entering an \circlearound{6}~\textit{Alternative Maneuver} infeasible. 
% The comparison of tasks derived from German legislation in Table \ref{TabTasksLegislation} shows that directly driving a vehicle from remote is currently prohibited in German legislation, and the \textit{Remote Driving} path during entering an \circlearound{6} \textit{Alternative Maneuver} is impossible. 

%\vfill

\begin{table}[]%[b]
\renewcommand{\arraystretch}{1.1} % Increase row spacing
\caption{Comparison of Defined Remote Operator Tasks \\and Tasks from German Legislation}
\label{TabTasksLegislation}
\centering
\begin{tabular}{|c|p{3.2cm}|p{3.8cm}|}
\cline{2-3}
\multicolumn{1}{c|}{} & \textbf{\rule{0pt}{2.5ex}Defined Tasks} & \textbf{\rule{0pt}{2.5ex}Tasks from Legislation} \\ 
\hline
\multirow{9}{*}{\rotatebox{90}{Driving-related \qquad \quad}} & \textbf{\rule{0pt}{2.5ex}Remote Assistance} & Maneuver Clearance \newline Maneuver Proposal\\ 
\cline{2-3}
& \textbf{\rule{0pt}{2.5ex}Remote Driving}& - \\
\cline{2-3}
&\textbf{\rule{0pt}{2.5ex}Remote Intervention} & \\
& \circlearound{2} De-/Activation & De-/Activation \\
& \circlearound{3} Automation Engagement & Reinitialisation\\
& \circlearound{5} Minimal Risk Maneuver & Minimal Risk Maneuver (by AV)%/ Emergency Stop 
\\
\cline{2-3}
& \textbf{\rule{0pt}{2.5ex}Remote Monitoring}& \\
& \circlearound{1} Start/End of Service & Removal \\
& \circlearound{4} Start/End of Monitoring & Assessment of Vehicle Condition \\
\hline
\multirow{2}{*}{\rotatebox{90}{\shortstack{Mission-\\related}\;\;}} & \textbf{\rule{0pt}{2.5ex}Communication} & Contacting Vehicle Occupants \newline Contact by Vehicle Occupants \\
\cline{2-3}
& \textbf{\rule{0pt}{2.5ex}Emergency Handling} & Data Documentation \newline Door Opening \\
\hline
\end{tabular}
\vspace{-0.35cm}
\end{table}

\begin{figure}[]
       \centering
       \includegraphics[scale=0.78]{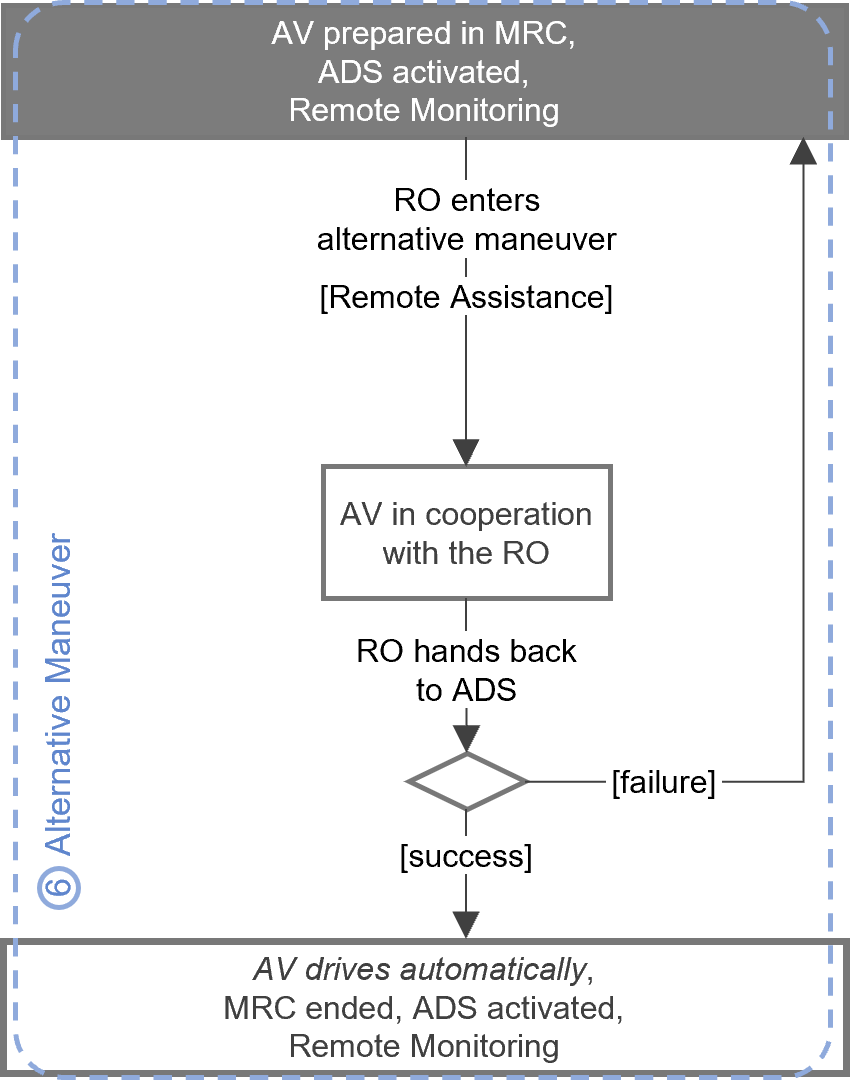}
       \vspace{-0.2cm}
       \caption{Adjusted excerpt of the state diagram for a solution compliant with German law~\cite{afgbv_general, stvg, eu-1426-2022}}
       \label{Excerpt}
       \vspace{-0.5cm}
    \end{figure}

The other tasks derived from legislation are considered in the developed state diagram or are assigned to the mission-related tasks defined in Section \ref{tasks} that can occur at any time during the remote operator's interaction with the \ac{av} or are taken over by the fleet manager. 
To be compliant with German legislation, the state diagram must be adjusted and remove the \textit{Remote Driving} transition (Fig. \ref{Excerpt}).

%%%%%%%%%%%%%%%%%%%%%%%%%%%%%%%%%%%%%%%%%%%%%%%%%%%%%%%%%%%%%%%%%%%%%%%%%

\section{DISCUSSION}
%The teleoperation framework, developed based on an extensive literature review of publicly accessible sources, provides for the first time a comprehensive state diagram of \ac{av} states in interaction with a remote operator, which can be operationalized globally for various applications. While teleoperation solutions are already implemented in the industry, for instance, by Waymo, Zoox, or Cruise,\todo[]{Quellen einfügen} no sources are detailing their implementation in a scientific context. 
Based on an extensive literature review, this work %the teleoperation framework 
presents the first comprehensive state diagram of an \ac{av} interacting with a remote operator, designed for global applications. While companies like Waymo~\cite{waymo}, or Zoox~\cite{zoox} %, or Cruise~\cite{cruise} 
use teleoperation, no scientific documentation of their methods exists. % no sources detail their methods in a scientific context. % like this work does. 
The proposed state diagram builds a basis for implementing and testing remote operator tasks. % within technical systems. 
% According to the current state, the framework 
% %and corresponding state diagram 
% shows no gaps or open issues. 
Furthermore, the framework supports improving existing systems and regulations, while identifying limitations and guiding further development.
%and to facilitate the framework's further development while identifying its limitations. 
As the framework is derived from literature and expert discussions, real-world operation tests must validate the state diagram. 
The testing should align with local regulations and may be guided by projects like SUNRISE \cite{sunrise2025}. 
Additionally, a certification body must verify legal compliance for on-road use. 
This requires a vehicle with appropriate sensors, actuators, and software. % for control and automation. 
While model- and simulation-based testing aids early development, real-world deployment may reveal additional challenges. 
% Implementation, deployment, and certification may reveal additional challenges. 
% This requires a vehicle equipped with appropriate sensors and actuators, as well as software for control and automated driving. 
% Implementation, deployment, and certification may uncover new challenges. 
% Model- or simulation-based testing can support early development, but many issues only arise in practical operation.
% However, the full range of challenges typically emerges only during real-world deployment.
%%%
%A real-world application allows the remote operator tasks to be supplemented if necessary and their logical sequence and process in the state diagram to be checked. 
%In practice, some remote operator tasks may also be automated to minimize human intervention as far as possible, as noted in Section \ref{framework}. 
Real-world testing will also show whether remote operator tasks can be automated to minimize human intervention, as noted in Section \ref{framework}. 
Remote operators are expected to resolve \acp{mrc}, though future \acp{av} may handle them independently with advanced intelligence and updated software. 
Nevertheless, the framework remains valid, applying only when \acp{av} cannot resolve \acp{mrc} on their own.
% It is assumed that the remote operator resolves an \ac{mrc}. % by engaging the automation or entering an alternative maneuver. 
% % In the future, \acp{av} may independently handle \acp{mrc} once the situation is clear, requiring advanced vehicle intelligence due to software or hardware updates. % and system trust. 
% Nevertheless, the general validity of the framework remains unchanged, with remote operator involvement limited to \acp{mrc} the \ac{av} cannot manage alone. %is only called upon to resolve \acp{mrc} that the \ac{av} cannot solve independently. 
Thereby, personnel requirements depend on fleet size, \ac{mrm} frequency, %with which an \ac{mrm} is triggered, 
and how to resolve the~\ac{mrc}, % the remote operator must resolve it, %requires resolution by a remote operator. 
e.g., \textit{Automation Engagement} requires less effort than \textit{Remote Driving}.
% We consider it reasonable for the \ac{mrc} to always be resolved by the remote operator by engaging the automation or entering an alternative maneuver. However, in the future, the \ac{av} may independently resolve the \ac{mrc} once the situation is clarified, requiring high vehicle intelligence and trust in the system. 
In addition, parameters such as when an \ac{av} requests a remote operator and the required response time must be defined and tested in practice. 
Furthermore, we assume that an \ac{mrm} is always performed safely. Otherwise, an emergency stop would be necessary %for the remote operator 
to override the \ac{ads} and trigger an immediate halt. %, e.g., during the \ac{av} attempts to reach an \ac{mrc}. 
Ultimately, the actual remote operator and \ac{av} interaction in the dynamic driving task is simplified %by the remote driving and assistance paths 
in the Alternative Maneuver block. 
Particularly, these different interaction possibilities, known as teleoperation concepts~\cite{majstorovic2022survey}, and the associated user interfaces require further research to enable safe teleoperation. %cooperation with the \ac{av} or remote driving. 

Developed in line with European legislation \cite{eu-1426-2022}, the framework appears transferable to other EU countries. However, applicability to other markets and domains must be assessed. % case by case. 
% Since the framework is also developed following European law \cite{eu-1426-2022}, its applicability to other EU countries seems feasible. 
% However, the transferability to other target markets and application areas requires case-specific evaluation. %still needs to be evaluated. 
Thereby, current regulations leave room for interpretation, while some countries lacking any teleoperation-specific law. 
% However, this work does not address the question of 
Liability for the consequences of the vehicle's actions is also not addressed in this work. 
Instead, the focus is on who performs the  %but solely identifies the party responsible for performing the 
dynamic driving task.

% Also, the process of handing over a successful teleoperation back to automation is still unclear. 
%Therefore, significant research is still needed to implement a control center for an \ac{av} fleet. 

%%%%%%%%%%%%%%%%%%%%%%%%%%%%%%%%%%%%%%%%%%%%%%%%%%%%%%%%%%%%%%%%%%%%%%%

\section{CONCLUSION}

Tasks in a control center for an \ac{av} fleet were gathered from literature and assigned to the specified roles: Remote Operator and Fleet Manager. Focusing on the driving-related tasks, the remote operator tasks were sequenced for an assumed \ac{av} service procedure, and a state diagram was developed. This state diagram illustrates the interaction between the remote operator and the \ac{av}, aiding the implementation of a technical teleoperation system. To assess the framework's applicability, the state diagram was adopted to German legislation. %by avoiding the prohibited Remote Driving path. 
The next step is to implement and test the created system in real-world conditions to verify its effectiveness in resolving diverse \ac{av} disengagements. 
%The next step is to implement and test the created system in the real world to verify whether this teleoperation framework can effectively resolve a high variety of \ac{av} disengagements.

\section*{ACKNOWLEDGMENT}
%As first author, Maria-Magdalena Wolf was responsible for clustering and defining the Teleoperation Control Center tasks and creating the corresponding state diagram. 
As the first author, Maria-Magdalena Wolf initiated the idea of this paper and contributed essentially to its conception and content. 
%defined and clustered the Teleoperation Control Center tasks and developed the state diagram. 
Niklas Krauss contributed to the literature research and content of this paper. 
%He was also involved in the critical discussion and finalization of the content.  
Arwed Schmidt and Frank Diermeyer made essential contributions to the conception of the research project and revised the paper critically. %for important intellectual content. 
%They gave final approval for the version to be published and agree to all aspects of the work. As guarantors, they accept responsibility for the overall integrity of the paper.
The publication is based on the funded research project SAFESTREAM, consisting of the partners EasyMile, P3, T-Systems, TÜV~Rheinland, the Technical~University~of~Munich, the City~Monheim~am~Rhein, and the District of Kelheim supported in legal discussions by Professor Ensthaler.
% This work was supported by the Federal Ministry of Economic Affairs and Climate Actions of Germany (BMWK) within the project SAFESTREAM (FKZ 01ME21007B), consisting of the partners EasyMile, P3, T-Systems, TÜV~Rheinland, the Technical~University~of~Munich, the City~Monheim~am~Rhein, and the District of Kelheim supported in legal discussions by Professor Ensthaler.
%The project consortium has been supported in continuous legal discussions by Professor Ensthaler.
Special thanks go to Andreas Schimpe for his initial input on the state diagram and to David Brecht for the discussions. %critical discussions

%%%%%%%%%%%%%%%%%%%%%%%%%%%%%%%%%%%%%%%%%%%%%%%%%%%%%%%%%%%%%%%%%%%%%%%%%%%%%%%%

%References are important to the reader; therefore, each citation must be complete and correct. If at all possible, references should be commonly available publications.

\bibliographystyle{IEEEtran}
\bibliography{bibliography}

\end{document}